\documentclass{article}
\usepackage[utf8]{inputenc}

\usepackage{biblatex}
\usepackage{authblk}

\usepackage{graphicx}
\graphicspath{ {./images/} }

\title{Multiclass Language Identification using Deep Learning on Spectral Images of Audio Signals}
\author[1]{Shauna Revay}
\author[1]{Matthew Teschke}
\affil[1]{Novetta}

\begin{document}

\maketitle

\section{Introduction}

Recently, voice assistants have become a staple in the flagship products of many big technology companies such as Google, Apple, Amazon, and Microsoft. One challenge for voice assistant products is that the language that a speaker is using needs to be preset. To improve user experience on this and similar tasks such as automated speech detection or speech to text transcription, automatic language detection is a necessary first step.

The technique described in this paper, language identification for audio spectrograms (LIFAS), uses spectrograms of raw audio signals as input to a convolutional neural network (CNN) to be used for language identification. One benefit of this process is that it requires minimal pre-processing. In fact, only the raw audio signals are input into the neural network, with the spectrograms generated as each batch is input to the network during training. Another benefit is that the technique can utilize short audio segments (approximately 4 seconds) for effective classification, necessary for voice assistants that need to identify language as soon as a speaker begins to talk. 

LIFAS binary language classification had an accuracy of 97\%, and multi-class classification with six languages had an accuracy of 89\%.  

\section{Background}

Finding a dataset of audio clips in various languages sufficiently large for training a network was an initial challenge for this task. Many datasets of this type are not open sourced \cite{mozilla}.  VoxForge \cite{voxforge}, an open-source corpus that consists of user-submitted audio clips in various languages, is the source of data used in this paper. 

Previous work in this area used deep networks as feature extractors, but did not use the networks themselves to classify the languages \cite{conference, unified}. LIFAS removes any feature extraction performed outside of the network. The network is fed a raw audio signal, and the spectrogram of the data is passed to the neural network during training. The last layer of the network outputs a vector of probabilities with one prediction per language. Thus, the whole process from raw audio signal to prediction of language is performed automatically by the neural network. 

In \cite{lstmpaper}, a CNN was combined with a long short-term memory (LSTM) network to classify language using spectrograms generated from audio. The network presented in \cite{lstmpaper} classified 4 languages using 10-second audio clips for training \cite{blog}, while LIFAS achieves similar performance for 6 languages using 4-second audio clips. This demonstrates the robustness of the architecture and its improvement upon earlier techniques.

\subsection{Residual and Convolutional Neural Networks}
CNNs have been shown to give state of the art results for image classification and a variety of other tasks. As neural networks using back propagation were constructed to be deeper, with more layers, they ran into the problem of vanishing gradient \cite{gradient}. A network updates its weights based on the partial derivatives of the error function from the previous layers. Many times, the derivatives can become very small and the weight updates become insignificant. This can lead to a degradation in performance.

One way to mitigate this problem is the use of Residual Neural Networks (ResNets \cite{resnet}). ResNets utilize skip connections in layers which connects two non-adjacent layers. ResNets have shown state-of-the-art performance on image recognition tasks, which makes them a natural choice for a network architecture for this task \cite{imageresidual}.

\subsection{Spectrogram Generation}

A spectrogram is an image representation of the frequencies present in a signal over time. The frequency spectrum of a signal can be generated from a time series signal using a Fourier Transform. 

In practice, the Fast Fourier Transform (FFT) can be applied to a section of the time series data to calculate the magnitude of the frequency spectrum for a fixed moment in time. This will correspond to a line in the spectrogram. The time series data is then windowed, usually in overlapping chunks, and the FFT data is strung together to form the spectrogram image which allows us to see how the frequencies change over time. 

Since we were generating spectrograms on audio data, the data was converted to the mel scale, generating "melspectrograms". These images will be referred to as simply "spectrograms" for the duration of this paper. The conversion from $f$ hertz to $m$ mels that we use is given by, 

$$m = 2595 \log_{10} \left( 1 + \frac{f}{700} \right).$$

\begin{figure}
    \centering
    \includegraphics[width=\textwidth]{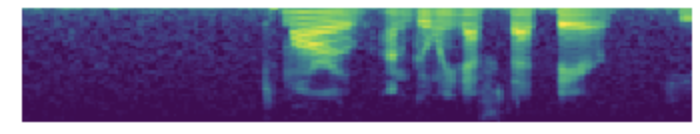}
    \caption{Spectrogram generated from an English audio file.}
    \label{spec}
\end{figure}

An example of a spectrogram generated by an English data transmission is shown in figure \ref{spec}.

\section{Data Preparation}

Audio data was collected from VoxForge \cite{voxforge}. Each audio signal was sampled at a rate of 16kHz and cut down to be 60,000 samples long. In this context, a sample refers to the number of data points in the audio clip. This equates to 3.75 seconds of audio. The audio files were saved as WAV files and loaded into Python using the librosa library and a sample rate of 16kHz.

Each audio file of 60,000 samples was saved separately and is referred to as a clip. The training set consisted of 5,000 clips per language, and the validation set consisted of 2,000 clips per language. 

Audio clips were gathered in English, Spanish, French, German, Russian, and Italian. Speakers had various accents and were of different genders. The same speakers may be speaking in more than one clip, but there was no cross contamination in the training and validation sets. 

Spectrograms were generated using parameters similar to the process discussed in \cite{audioblog} which used a frequency spectrum of 20Hz to 8,000Hz and 40 frequency bins. Each FFT was computed on a window of 1024 samples. No other pre-processing was done on the audio files. Spectrograms were generated on-the-fly on a per-batch basis while the network was running (i.e. spectrograms were not saved to disk).

\section{Network}

We utilized the fast.ai \cite{fastai} deep learning library built on PyTorch \cite{pytorch}. The network used was a pretrained Resnet50. The spectrograms were generated on a per-batch basis, with a batch size of 64 images. Each image was $432 \times 288$ pixels in size.

During training, the 1-cycle-policy described in \cite{leslie} was used. In this process, the learning rate is gradually increased and then decreased in a linear fashion during one cycle \cite{onecycleblog}.  The learning rate finder within the fast.ai library was first used to determine the maximum learning rate to be used in the 1-cycle training of the network. The maximum learning rate was then set to be $1 \times 10^{-2}$. The learning rate increases until it hits the maximum learning rate, and then it gradually decreases again. The length of the cycle was set to be 8 epochs, meaning that throughout the cycle 8 epochs are evaluated.

\section{Experiments}

\subsection{Binary Classification with Varying Number of Samples}
Binary classification was performed on two languages using clips of 60,000 samples. English and Russian were chosen to use for training and validation. To test the impact of the number of samples on classification while keeping the sample rate constant, binary classification was also performed on clips of 100,000 samples. 

\subsection{Multiple Language Classification}
For each language (English, Spanish, German, French, Russian, and Italian), 5,000 clips were placed in the training set. Each clip was 60,000 samples in length. 2,000 clips per language were placed in the validation set, and no speakers or clips appeared in both the training and validation sets.

\section{Results}
Accuracy was calculated for both binary classification and multiclass classification as: $$Accuracy = \frac{Number \; of \; Correct \; Predictions}{Total \;Number \;of \; Predictions}.$$ 
LIFAS binary classification accuracy for Russian and English clips of length 60,000 samples was 94\%. In comparison, LIFAS binary classification accuracy on the clips of 100,000 samples was 97 \%. The accuracy totals given in the experiments above are calculated on the total number of clips in the validation set. The accuracy can also be broken up into accuracy for English clips, or accuracy for Russian clips, where there was essentially no difference in the accuracy for English clips and the accuracy for Russian clips. 

To confirm that the network performance was not dependent on English and Russian language data, binary classification was tested on other languages with little to no impact on validation accuracy. 

LIFAS accuracy for the multi-class network with six languages was 89 \%. These results were based on clips of 60,000 samples since a sufficient number of longer clips were unavailable. Results from the 100,000 sample clips in the binary classification model suggest performance could be improved in the multi-class setting with longer clips. 

\begin{figure}
    \centering
    \includegraphics[width=0.8\textwidth]{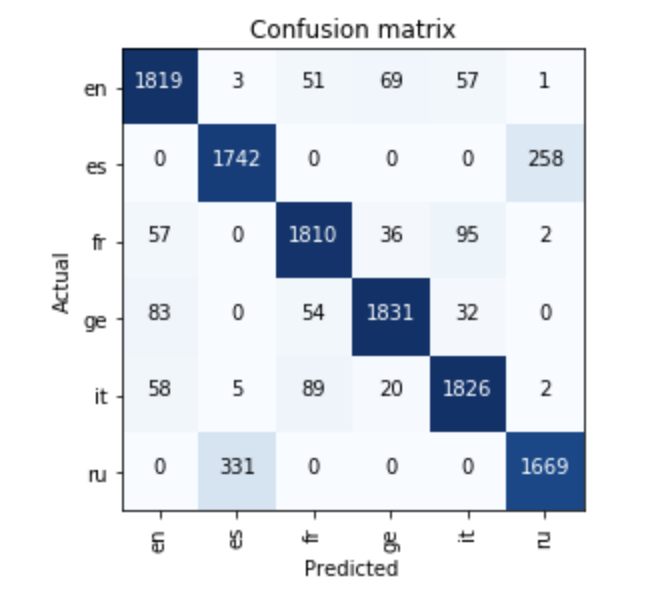}
    \caption{The confusion matrix for the multiclass language identification problem.}
    \label{confusion}
\end{figure}

The confusion matrix for the multi-class classification is shown in figure \ref{confusion}.

\section{Discussion and Limitations}

Notably, the highest rate of false negative classifications came when Spanish clips were classified as Russian, and when Russian clips were classified as Spanish. Additionally, almost no other language is misclassified as Russian or Spanish. One hypothesis for this observation is the fact that Russian is the only Slavic language in the training set. Therefore, the network may be performing some thresholding at one layer that separates Russian from other languages, and by chance Spanish clips are near the threshold.  

One limitation in our findings is that all of the data came from the same dataset. Since audio formats can have a wide variety of parameters such as bit rate, sampling rate, and bits per sample, we would expect clips from other datasets collected in different formats to potentially confuse the network. There is potential for this drawback to be overcome assuming appropriate pre-processing steps were taken for the audio signals so that the spectrograms did not contain artifacts from the dataset itself. This is a problem that should be explored as more data becomes available. 

\section{Conclusion}

This work shows the viability of using deep network architectures commonly used for image classification in identifying languages from images generated from audio data. Robust performance can be accomplished using relatively short files with minimal pre-processing. We believe that this model can be extended to classify more languages so long as sufficient, representative training and validation data is available. A next step in testing the robustness of this model would be to include test data from additional (e.g. non-VoxForge) datasets. 

Additionally, we would want the network to be performant on environments with varying levels of noise. VoxForge data is all user submitted audio clips, so the noise profiles of the clips vary, but more regimented tests should be done to see how robust the network is to different measured levels of noise. Simulated additive white Gaussian noise could be added to the training data to simulate low quality audio, but still might not fully mimic the effect of background noise such as car horns, clanging pots, or multiple speakers in a real life environment. 

Another way to potentially increase the robustness of the model would be to implement SpecAugment \cite{specaugment} which is a method that distorts spectrogram images in order to help overfitting and increase performance of networks by feeding in deliberately corrupted images. This may help to add scalability and robustness to the network, assuming the spectral distortions generated in SpecAugment accurately represent distortions in audio signals observed in the real world. 
\printbibliography

\end{document}